\documentclass[12pt,a4paper]{article}
\usepackage[dvips]{graphics}
\usepackage[dvips]{graphicx}
\usepackage{enumerate}
\usepackage{psfrag}
\usepackage{amsmath}
\usepackage{amssymb}
\usepackage{epsfig}
\usepackage{rotating}
\usepackage{mathrsfs}
\usepackage{accents}

\def\m3#1#2#3{($\mathbf{#1}, \mathbf{#2}, \mathbf{#3}$)}
\newcommand{\mathsym}[1]{{}}

\newcommand{\newc}{\newcommand}
\newc{\ba}{\begin{eqnarray}}
 \newc{\ea}{\end{eqnarray}}
\oddsidemargin \evensidemargin
\newcommand{\beq}{\begin{equation}}
\newcommand{\eeq}{\end{equation}}

\newcommand{\RaiseBrace}[1]{\raise1.5pt\hbox{$\displaystyle#1$}}

\catcode`\@=11
\def\@normalsize{\@setsize\normalsize{15pt}\xiipt\@xiipt
\abovedisplayskip 14pt plus3pt minus3pt%
\belowdisplayskip \abovedisplayskip
\abovedisplayshortskip  \z@ plus3pt%
\belowdisplayshortskip  7pt plus3.5pt minus0pt}
\def\small{\@setsize\small{13.6pt}\xipt\@xipt
\abovedisplayskip 13pt plus3pt minus3pt%
\belowdisplayskip \abovedisplayskip
\abovedisplayshortskip  \z@ plus3pt%
\belowdisplayshortskip  7pt plus3.5pt minus0pt
\def\@listi{\parsep 4.5pt plus 2pt minus 1pt
            \itemsep \parsep
            \topsep 9pt plus 3pt minus 3pt}}
\def\underline#1{\relax\ifmmode\@@underline#1\else
        $\@@underline{\hbox{#1}}$\relax\fi}
\@twosidetrue \relax \catcode`@=12
\def\ps@headings{\def\@oddfoot{}\def\@evenfoot{}
\def\@oddhead{\hbox{}\hfill
        \makebox[.5\textwidth]{\raggedright\ignorespaces --\thepage{}--
        \hfill }}
\def\@evenhead{\@oddhead}
\def\subsectionmark##1{\markboth{##1}{}}
} \ps@headings \catcode`\@=12 \relax

\begin{document}

 \begin{titlepage}

\begin{centering}
\vspace*{1cm}
{\Large \bf Neutrino and Multidimensional Theories of}\\[4mm]
{\Large \bf Fundamental Interactions}\\
\vspace{2.5cm} {\bf Psallidas Andreas}\footnote{\noindent e-mail~:
me00604@cc.uoi.gr . A full version of the dissertation in Greek can be found in~: http://users.uoi.gr/neutrino/PHD.pdf}\\[15mm]
{\it Physics Department, Section of Theoretical Physics,\\[2mm]
University of Ioannina, Greece}\\[35mm]

%\vspace*{0.5in} \centerline{\thicklines \epsfysize=5cm \epsfbox{f1.eps}}
%\centerline{\psfig{figure=f1.ps}}

\vspace*{0.6in}
{\it A Dissertation in Candidacy for the Degree of Doctor of Philosophy}\\[55mm]

Thesis Advisor: G. K. Leontaris \\[5mm] \centerline{{October 2006}}

\end{centering}

 \end{titlepage}

 \newpage

%\begin{center}
% \begin{figure}
%\includegraphics[width=1.0\textwidth]{logotupo.eps}
%\end{figure}
% \end{center}

  \begin{center}
  {\bf Abstract}
 \end{center}
 \vspace{0.6cm}
 In this thesis, the possibility of interpreting the solar and
atmospheric neutrino data within the context of theoretical models is being
explored. In particular, the implications of the Minimal Supersymmetric Standard
Model augmented by a single $U(1)$ anomalous family symmetry for neutrino masses
and mixing angles are investigated. The family symmetry is spontaneously broken
by non--zero vacuum expectation values of a pair of singlet fields. The symmetry
retains a dimension-five operator which provides Majorana masses for left-handed
neutrino states. Assuming symmetric lepton mass matrices, the model can account
for atmospheric data and predicts $\theta_{13}=0$. It is also shown how under
certain assumptions the model can be compatible with all recent experimental
data.
%secondary effects possibly arising from additional singlet(s) or some
%alternative mechanism, as supersymmetry breaking,
Motivated by the above results and the fact that string theory models predict in
their spectrum a large number of neutral singlet fields, the effect of
introducing  a proper second singlet pair is studied resulting to additional mass
entries in the previous model. Consequently, the general form of the charged
lepton and neutrino mass matrices is derived when two different pairs of singlet
Higgs fields develop non--zero vacuum expectation values and the resulting
neutrino textures are related to approximate lepton flavor symmetries. A
numerical analysis for one particular case is performed and solutions are
obtained for masses and mixing angles, consistent with  experimental data. Also,
renormalization effects on neutrino masses and mixing angles in a supersymmetric
string-inspired SU$(4)\times$SU$(2)_L\times$SU$(2)_R\times$U$(1)_X$ model are
discussed, with matter in fundamental and antisymmetric tensor representations
and singlet Higgs fields
 charged under the anomalous $U(1)_X$ family symmetry. The quark, lepton and neutrino Yukawa
 matrices are distinguished by different Clebsch-Gordan coefficients.
 The presence of a second $U(1)_X$ breaking singlet with fractional charge allows a more realistic,
 hierarchical light neutrino mass spectrum with bi-large mixing. By numerical investigation a
 region is found in the model's parameter space where the neutrino mass-squared
 differences and mixing angles at low energy are
 consistent with experimental data. Lastly, D-brane inspired models are studied with $U(3)\times U(2)\times U(1)^N$ gauge
symmetry in the context of split supersymmetry. Configurations with one, two and
three ($N=1,2,3$) abelian factors are considered and all hypercharge embeddings
are derived which imply the realistic particle content of the Standard Model with
the addition of the right-handed neutrino. Then, the implications of split
supersymmetry on the magnitude of the string scale, the gauge coupling evolution
and the third family fermion mass relations are analyzed. Gauge coupling
relations are considered which may arise in parallel as well as intersecting
brane scenarios and the various models are classified according to their
predictions for the magnitude of the string scale and the low energy
implications. In the parallel brane scenario where the $U(1)$ branes are
superposed to $U(2)$ or $U(3)$ brane stacks, varying the split susy scale in a
wide range, three distinct cases of models are found predicting a high,
intermediate and low string scale, $M_S\sim 10^{16}$ GeV, $M_S\sim 10^{7}$ GeV
and $M_S\sim 10^{4}$ GeV respectively. Furthermore, in the intermediate string
scale model the low energy ratio $m_b/m_{\tau}$ is compatible with $b-\tau$
Yukawa unification at the string scale. Moreover, a similar analysis is performed
for arbitrary gauge coupling relations at $M_S$ corresponding to possible
intersecting brane models. Cases that predict a string scale of the order $M_S\ge
10^{14}$ GeV are explored which accommodate a right-handed neutrino mass of the
same order so that a see-saw type light left-handed neutrino component is
obtained in the sub-eV range as required by experimental and cosmological data.
Finally, a discussion is devoted for the life-time of the gluino.\\[10mm]  \
 \vspace*{5mm}
\centerline{\Large \bf Publications} \

\noindent
 [1] D.~V.~Gioutsos, G.~K.~Leontaris and A.~Psallidas,
   "D-brane standard model variants and split supersymmetry: Unification  and
  fermion mass predictions",
 \textit{Physical Review \textbf{D} 74}, 075007, 2006, hep-ph/0605187.
 \\[2mm]
\noindent
 [2] T.~Dent, G.~K.~Leontaris, A.~Psallidas and J.~Rizos,
   "Renormalization effects on neutrino masses and mixing in a  string-inspired
  SU(4) x SU(2)$_L$ x SU(2)$_R$ x U(1)$_X$ model", \textit{submitted to Physical Review \textbf{D}}, hep-ph/0603228.\\[2mm]
\noindent
 [3] G.~K.~Leontaris, A.~Psallidas and N.~D.~Vlachos,
 "Inverted neutrino mass hierarchies from U(1) symmetries", \textit{accepted in International Journal of
 Modern Physics A}, hep-ph/0511327.\\[2mm]
\noindent
  [4] G.~K.~Leontaris, J.~Rizos and A.~Psallidas, "Majorana
neutrino masses from anomalous U(1) symmetries", \textit{Physics
  Letters  \textbf{B}597}: 182-191, 2004, hep-ph/0404129.\\[2mm]
\noindent

 \newpage

\centerline{\Large\bf Contents}
\vspace*{8mm} \noindent
{Preface} \dotfill 14\\[5mm]
{\bf 1. Introduction to the physics of neutrinos} \dotfill 19\\[3mm]
\indent 1.1 Introduction \dotfill {\rm 19}\\[2mm]
\indent 1.2 Solar and atmospheric neutrinos \dotfill {\rm 20}\\[2mm]
\indent 1.3 Neutrino detection \dotfill {\rm 22}\\[2mm]
\indent 1.4 Neutrino oscillations \dotfill {\rm 24}\\[2mm]
\indent \indent 1.4.1 Neutrino oscillations in vacuum \dotfill {\rm 25}\\[2mm]
\indent \indent 1.4.2 Neutrino oscillations in matter \dotfill {\rm 29}\\[2mm]
\indent 1.5 Fermion masses \dotfill {\rm 34}\\[2mm]
\indent \indent 1.5.1 Fermion masses in the Standard Model \dotfill {\rm 35}\\[2mm]
\indent \indent 1.5.2 Fermion masses beyond the Standard Model \dotfill {\rm 36}\\[2mm]
\indent \indent 1.5.3 The see-saw mechanism in the Standard Model \dotfill {\rm 37}\\[2mm]
\indent 1.6 Experimental data and neutrino \dotfill {\rm 39}\\[5mm]
{\bf 2. Introduction to supersymmetry} \dotfill {\rm 43}\\[3mm]
\indent 2.1 Introduction \dotfill {\rm 43}\\[2mm]
\indent 2.2 The supersymmetric algebra \dotfill {\rm 43}\\[2mm]
\indent 2.3 The formalism of superspace and superfields \dotfill {\rm 45}\\[2mm]
\indent 2.4 Chiral and vector superfields \dotfill {\rm 48}\\[2mm]
\indent \indent 2.4.1 Chiral superfields \dotfill {\rm 48}\\[2mm]
\indent \indent 2.4.2 Vector superfields \dotfill {\rm 50}\\[2mm]
\indent 2.5 Supersymmetric lagrangians \dotfill {\rm 51}\\[2mm]
\indent 2.6 Supersymmetry breaking \dotfill {\rm 55}\\[5mm]
{\bf 3. The Minimal Supersymmetric Standard Model and its extensions} \dotfill
{\rm 57}\\[3mm]
\indent  3.1 Introduction \dotfill {\rm 57}\\[2mm]
\indent  3.2 Description of the model \dotfill {\rm 58}\\[2mm]
\indent  3.3 Spontaneous electroweak symmetry breaking in the MSSM \dotfill {\rm 60}\\[2mm]
\indent  3.4 The spectrum of the MSSM \dotfill {\rm 62}\\[2mm]
\indent  3.5 The Froggatt-Nielsen mechanism and its generalizations \dotfill {\rm 64}\\[2mm]
\indent  3.6 Anomalies \dotfill {\rm 67}\\[5mm]
{\bf 4. D-Branes and the effective field theory approximation} \dotfill {\rm 71}\\[3mm]
\indent  4.1 Introduction \dotfill {\rm 71}\\[2mm]
\indent  4.2 Open strings in Dp-Branes \dotfill {\rm 71}\\[2mm]
\indent  4.3 Open strings between Dp-Branes and Dq-Branes \dotfill {\rm 74}\\[2mm]
\indent \indent 4.3.1 Dp-Branes of same dimension \dotfill {\rm 74}\\[2mm]
\indent \indent 4.3.2 Dp-Branes of different dimension \dotfill {\rm 77}\\[2mm]
\indent  4.4 Intersecting branes \dotfill {\rm 78}\\[2mm]
\indent  4.5 The energy scale of the string and the effective theory \dotfill {\rm 79}\\[5mm]
\noindent {\bf 5.  Extending the  MSSM with an anomalous $U(1)_X $ symmetry} \dotfill {\rm 81}\\[2mm]
\indent  5.1 Introduction \dotfill {\rm 81}\\[2mm]
\indent  5.2 Motivation \dotfill {\rm 82}\\[2mm]
\indent  5.3 Description of the model \dotfill {\rm 83}\\[2mm]
\indent  5.4 Neutrino masses and the mixing angles \dotfill {\rm 87}\\[2mm]
\indent  5.5 Approximation with 2 pairs of singlet fields \dotfill {\rm 92}\\[2mm]
\indent \indent 5.5.1 Bounds to the mixing angles and masses of the neutrinos \dotfill {\rm 93}\\[2mm]
\indent \indent 5.5.2 The addition of the second pair of singlet fields \dotfill {\rm 97}\\[2mm]
\indent \indent 5.5.3 Numerical analysis \dotfill {\rm 100}\\[5mm]
{\bf 6. Pati-Salam models} \dotfill {\rm 107}\\[2mm]
\indent  6.1 Introduction \dotfill {\rm 107}\\[2mm]
\indent  6.2 The supersymmetric $SU(4) \times SU(2)_L \times SU(2)_R$ model \dotfill {\rm 108}\\[2mm]
\indent  6.3 Renormalization Group Equations \dotfill {\rm 111}\\[2mm]
\indent  6.4 The supersymmetric $SU(4)\times SU(2)_L\times SU(2)_R\times U(1)_X$
model \dotfill {\rm 113}\\[2mm]
\indent \indent 6.4.1 The mass matrices of the fermions \dotfill {\rm 117}\\[2mm]
\indent \indent 6.4.2 Evolution of masses and mixing angles \dotfill {\rm 122}\\[5mm]
{\bf 7. Supersymmetric and non supersymmetric configurations of D-Branes} \dotfill {\rm 131}\\[2mm]
\indent  7.1 Introduction \dotfill {\rm 131}\\[2mm]
\indent  7.2 The Standard Model from D-Branes \dotfill {\rm 133}\\[2mm]
\indent  7.3 String scale and Yukawa couplings in supersymmetric and \\[1mm] \indent non supersymmetric configuration \dotfill {\rm 143}\\[2mm]
\indent  7.4 Split supersymmetry \dotfill {\rm 149}\\[2mm]
\indent  7.5 Unification and the string scale \dotfill {\rm 150}\\[2mm]
\indent \indent 7.5.1 Parallel D-Branes \dotfill {\rm 155}\\[2mm]
\indent \indent 7.5.2 Intersecting branes \dotfill {\rm 158}\\[2mm]
\indent \indent 7.5.3 b-$\tau$ unification \dotfill {\rm 163}\\[2mm]
\indent \indent 7.5.4 Masses, gaugino couplings and the lifetime of the gluino \dotfill {\rm 166}\\[5mm]
\bf Conclusions \dotfill {\rm 174}\\[5mm]
\bf Appendixes \dotfill {\rm 179}\\[5mm]
{\bf  A. The mass matrices of the fermions and the mixing angles in the \\
supersymmetric $SU(3)_C \times SU(2)_L \times U(1)_Y \times U(1)_X$}\dotfill {\rm 179}\\[5mm]
\indent  A.1 Quark sector \dotfill {\rm 179}\\[2mm]
\indent  A.2 Lepton sector \dotfill {\rm 184}\\[5mm]
{\bf B. Normalization of hypercharge } \dotfill {\rm 193}\\[5mm]
{\bf  C. Renormalization Group Equations} \dotfill {\rm 197}\\[5mm]
\indent  C.1 The extension of the MSSM with heavy singlet fields \dotfill {\rm 197}\\[2mm]
\indent  C.2 Split supersymmetry \dotfill {\rm 199}\\[5mm]
{\bf D. Program} \dotfill {\rm 203}\\[5mm]
\bf Bibliography \dotfill {\rm 209}\\[5mm]
\bf Terminology \dotfill {\rm 217}

\end{document}